\documentclass{appolb}
\usepackage{epsfig}

\begin{document}

\title{Noise Enhanced Stability
\thanks{Presented at the XVI Marian Smoluchowski Symposium on Statistical Physics, \ \\
Zakopane, Poland, September 6-11, 2003.}}

\author{Bernardo Spagnolo
\address{INFM-Group of Interdisciplinary Physics and  Dipartimento di
Fisica e Tecnologie Relative, Universit\`a di Palermo, Viale delle
Scienze - 90128 Palermo, Italy} \and Nikolay V. Agudov, Alexander
A. Dubkov
\address{Radiophysics Department, Nizhni
Novgorod State University, 23 Gagarin Ave., 603950 Nizhni
Novgorod, Russia}} \maketitle

\begin{abstract}
The noise can stabilize a fluctuating or a periodically driven
metastable state in such a way that the system remains in this
state for a longer time than in the absence of white noise. This
is the noise enhanced stability phenomenon, observed
experimentally and numerically in different physical systems.
After shortly reviewing all the physical systems where the
phenomenon was observed, the theoretical approaches used to
explain the effect are presented. Specifically the conditions to
observe the effect: (a) in systems with periodical driving force,
and (b) in random dichotomous driving force, are discussed. In
case (b) we review the analytical results concerning the mean
first passage time and the nonlinear relaxation time as a function
of the white noise intensity, the parameters of the potential
barrier, and of the dichotomous noise.
\end{abstract}

\section{Introduction}

The escape from metastable states continues to attract increasing
interest since the Kramers' seminal paper \cite{Kra}. It occurs in
a wide variety of natural systems such as chemical systems, spin
systems, quantum liquids, polymers and in problems of transport in
complex systems, such as glasses and proteins. Specifically the
noise activated escape in systems with metastable states and
fluctuating barriers is important to describe the dynamics of
complex nonequilibrium systems such as the molecular dissociation
in strongly coupled chemical systems, electron transport in a
quantum double-well structure, crystal growth, glasses,
microstructures, lasers, Josephson junction devices, ratchet
models, migration of ligands in proteins and biological systems
\cite{Han}-\cite{Rei}. A common peculiarity of all these systems
is that are open systems with internal nonlinear dynamics and
interacting with a noisy environment, which is responsible for
noise induced phenomena. In these complex nonstationary
nonequilibrium systems the continuous time-translation symmetry is
broken, in contrast to the phenomenon of stochastic resonance
characterized by deterministic barrier modulations \cite{Gam}.\\
\indent Noise activated escape from a metastable state with
oscillating or fluctuating barriers has recently attracted
increasing attention \cite{Sme}-\cite{Agu2}. In many situations
the system is driven away from thermal equilibrium by an
additional periodical driving force or by some external random
perturbations. While important from both fundamental and
application point of view, analytical progress in the theory of
oscillating barrier crossing is rather difficult. In the weak
noise regime an interesting phenomenon appears: the enhancement of
stability by thermal noise in systems with a metastable state and
a periodically driven potential \cite{Day,Man}. This noise
enhanced stability phenomenon (NES) was observed experimentally
and numerically in various physical systems
\cite{Agu1}-\cite{Xie}. By varying the value of the thermal noise
intensity we can lengthen or shorten the mean lifetime of the
metastable state of our physical system. The enhancement of
stability implies that the system remains in the metastable state
for a longer time than in the absence of noise.\\
\indent The paper is organized as follows. In the first section we
shortly review
 all the physical systems where the effect was observed. The
theoretical approaches that we used to explain the effect are
presented in second and third sections. Specifically the
conditions for the NES effect: (a) with periodical driving force,
and (b) with random dichotomous driving force, are discussed. In
the final section we review for the case (b) the analytical
results concerning the mean first passage time (MFPT) and the
nonlinear relaxation time (NLRT) as a function of the white noise
intensity, the parameters of the potential barrier, and of the
dichotomous noise.

\section{Enhancement of Stability in Physical Systems}

The mean first passage time of a Brownian particle moving in
potential fields with metastable and unstable states normally
decreases with noise intensity growth according to the Kramers'
formula \cite{Kra}
\begin{equation}
\tau_k = A\cdot \e^{\Delta U/q} \label{kramer}
\end{equation}
or some universal scaling function of the system parameters
\cite{Col}. In equation (\ref{kramer}) $A$ is a pre-factor, which
depends on the curvature of the potential at the metastable state
and at the top of the potential barrier of height $\Delta U$, and
$2q$ is the noise intensity. However, the dependence of the MFPT
for unstable or oscillating metastable states, was revealed to
have resonance character with a nonmonotonic behavior as a
function of the noise intensity. This is the NES phenomenon: the
noise can modify the stability of the system. Under the action of
noise a system remains in the unstable or in the oscillating
metastable state for a longer time than in the deterministic case
and the escape time as a function of noise intensity has a
maximum. The NES phenomenon has been observed in different
physical systems, which we review in this section.\\
\indent Hirsch et al. \cite{Hir} first noted this nonmonotonic
dependence in studying the onset of intermittent chaotic behavior
for one-dimensional maps just before a tangent bifurcation. They
considered the effect of external noise on the regular path length
and obtained an interesting result: for some values of the system
parameters the average length of the laminar regions may be
enhanced by the presence of a given finite amount of noise. Hirsh
and coauthors considered a Langevin equation with the potential
corresponding to the unstable state and reduced the average path
length of the laminar regime problem to the mean first passage
time problem. They found that \emph{a small amount of noise
increase the average time of passage, contrary to what one might
have expected}. A simple model which exhibits the same phenomenon
was studied by Agudov and Malakhov \cite{Agu5}. They considered a
logistic map with a piecewise linear function and studied this
effect in detail.\\
\indent In a theoretical study of the transient dynamics of an
overdamped Brownian particle in a time dependent cubic potential
with metastable state (see Fig.~1), Dayan et al. \cite{Day} showed
numerically that the stability of the system is enhanced for a
wide choice of values of the control parameter.
\begin{figure}[htbp]
\begin{center}
\includegraphics[height=5cm,width=6cm]{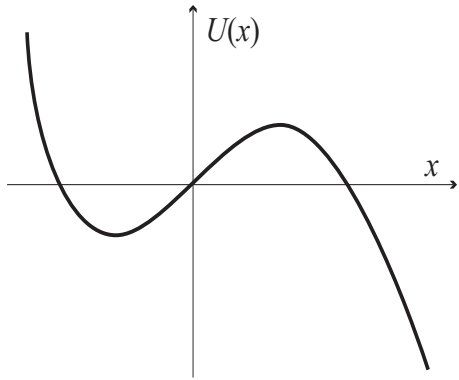}
\end{center}
\caption{\small \emph{Cubic potential with a metastable state.}}
\label{cubpot}
\end{figure}
The initial intuition of authors was that an increase in the
amplitude of the noise should decrease the mean escape time
because noise forces the particle to sample more of the available
space than without noise. However their simulation studies showed
that at an appropriately chosen frequency of the periodical
driving force \emph{the escape time increases when the intensity
of the noise increases}. This means that the noise can modify the
stability of the system in a counter-intuitive way. This effect
was named by Mantegna and Spagnolo \cite{Man} as Noise Enhanced
Stability (NES) and gives a nonmonotonic behavior of the average
escape time as a function of the noise intensity. The NES
phenomenon was experimentally observed by investigating the escape
time from a time modulated physical system: the tunnel diode
\cite{Man}. The system can be deterministically overall stable or
overall unstable. In the presence of noise the overall-stable
regime becomes metastable, the stability of the system is higher
for lower values of noise amplitude. In the overall-unstable
regime, however, a finite amount of \emph{noise increase the
stability} of the otherwise deterministically unstable system. A
related phenomenon was revealed and investigated by Agudov and
Malakhov in ref.~\cite{Agu3,Agu4} for different kinds of fixed
potential profiles. This is the noise delayed decay (NDD) of
unstable nonequilibrium states. Previous investigations showed
that noise accelerated the decay of any unstable state
\cite{Col,Cru}. Agudov and Malakhov by analyzing the influence of
the potential profile shape and initial conditions on the NDD
effect showed that the decay time of unstable states, under some
conditions, can be increased considerably by the external noise.
In other words, \emph{the external additive noise can delay the
decay of unstable states}.\\
\indent We note that the NDD and NES effects are two different
aspects of the same noise induced phenomenon occurring in
nonlinear physical systems, but  with some peculiarities. The NDD
effect concerns the delay of the decay of unstable nonequilibrium
initial states in fixed potential profiles \cite{Agu3,Agu4}. The
NES effect appears in potential profiles with metastable state in
the presence of a strong driving force. The dynamical regime is
characterized by the absence of the potential barrier for some
short time interval; that is, the system is
\emph{deterministically overall unstable} \cite{Man} and in this
time interval we have the same physical situation as for NDD
effect. After this time interval the Brownian particle, because of
the interplay between the noise and the time dependent driving
force, can return into the potential well and the mean life time
of the metastable state increases with the noise intensity, in
comparison with the dynamical life time \cite{Day,Man,Agu1}. When
we consider fixed potentials with \emph{metastable state} and with
\emph{initial unstable positions}, we can refer to both NDD or NES
effect. In this case we have a nonmonotonic behavior of the
average escape time as a function of noise intensity and a new
interesting dynamical regime, characterized by
 a \emph{divergency} of the average escape time with an exponential
 Kramers-like
behavior\cite{Agu1,Fia}.\\
\indent By investigating the influence of thermal fluctuations on
the superconductive state lifetime and the turn-on delay time for
a single Josephson element with high damping, Malakhov and
Pankratov \cite{Mal} found that fluctuations may both decrease and
increase the turn-on delay time. Specifically they found that for
low noise intensities and for current values greater than the
critical current, which characterize the onset of the resistive
state from a superconductive state, \emph{an increase of
fluctuations intensity causes increasing of the metastable state
lifetime}.\\
\indent Wackerbauer analyzed in detail the influence of dynamical
noise on switching processes in one-dimensional discontinuous maps
\cite{Wac}, namely the piecewise linear map, the Lorenz map and
the piecewise linear Lorenz map. The main result is that the
switching dynamics of all Lorenz-type maps is significantly
reduced by dynamical noise. This reduction is mainly caused by a
noise-induced escape of a typical trajectory into a less
frequently visited part of the attractor. This causes a
\emph{noise-induced stabilization, \ie an enhancement of the mean
passage time}. Several properties found in the noisy Lorenz system
are related to findings in the transient dynamics of a modulated
metastable system, which shows the NES phenomenon \cite{Man}.\\
\indent The mobility of an overdamped particle in a periodic
potential tilted by a constant external field and moving in a
medium with periodic friction coefficient shows noise induced
slowing down \cite{Mah}. For large values of the constant external
field, for which the potential barrier disappears, the mobility
decreases as the intensity of the thermal noise increases from
zero temperature, where one would have expected the particle to
become more mobile as the temperature is increased from zero. The
presence of \emph{noise slows down the motion of deterministically
overall unstable states} in an appropriate range of potential
parameters, contrary to what one might have expected. This is
somewhat akin to the phenomenon of noise enhanced stability of
unstable states \cite{Man}.\\
\indent The overdamped motion of a Brownian particle in an
asymmetric bistable fluctuating potential shows \emph{noise
induced stability} of the state which has most of the time the
higher energy \cite{Mie}. For intermediate fluctuation rates the
mean occupancy of minima with energy above the absolute minimum is
enhanced. The less stable minimum most of the time is metastable
and nevertheless it can be highly occupied.\\
\indent Yoshimoto showed in recent papers \cite{Yos} that one type
of \emph{noise-induced order}, in one-dimensional return map of
the Belousov-Zhabotinsky reaction, takes place in the intermittent
chaos, when the length of the laminar region was increased by the
noise. Finally Xie and Mei studied in ref.~\cite{Xie} the
transient properties of a bistable kinetic system driven by two
correlated noises: an additive noise and a multiplicative colored
noise. They found \emph{one-peak structure in the mean first
passage time} (MFPT) as a function of noise intensity for strongly
correlated noises. The peak grows highly as the correlation time
and the cross-correlation coefficient increase, which means that
the noise color causes the suppression effect of the escape rate
to become more pronounced, \ie the enhancement of the average
escape time with increasing noise intensity.

\section{Periodical Driving Force}

\subsection{Dichotomous Driving}

We consider the model of overdamped Brownian motion described by
the equation
\begin{equation}
\frac{dx}{dt}=-\frac{dU\left(x\right)}{dx}+F\left(
t\right)  +\xi\left(  t\right)  ,
\label{f-1}%
\end{equation}
where $\xi(t)$ is the white Gaussian noise with zero mean,
$\left\langle \xi(t)\xi (t+\tau)\right\rangle =2q\delta(\tau)$,
$F(t)$ is the dichotomous driving force and $U(x)$ is a piecewise
potential with the reflecting boundary and a metastable state at
$x=0$, maximum at $x=L$, and the absorbing boundary at $x=b$ (see
Fig. 2)
\begin{equation}
U\left(  x\right)  =\left\{
\begin{array}
[c]{cc}%
+\infty, & x<0\\ hx, & 0\leq x\leq L,\\ E-k\left(  x-L\right)  , &
L<x<b
\end{array}
\right. \label{f-2}%
\end{equation}
where $h>0,\;k>0$ and $E=hL$ is the height of the potential
barrier.

\begin{figure}[htbp]
\begin{center}
\includegraphics[width=9cm,height=9cm]{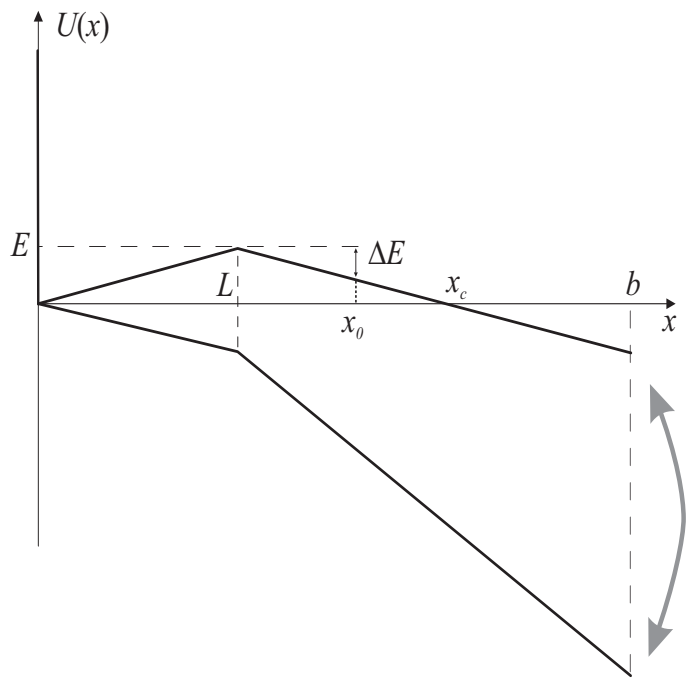}
\end{center}
\caption{\small \emph{The piecewise linear potential with
metastable and unstable states.}} \label{piecewise}
\end{figure}
For \emph{fixed potential} $(F(t)=0)$, when the particle is within
the potential well the decay time of metastable state increases
exponentially in the limit $q\rightarrow0$, according to Kramers'
formula (\ref{kramer}). The MFPT from initial position $x_0$ to
boundary $b$ is

\begin{equation}
\tau(x_0,b) = \frac{1}{q}\int_{x_0}^{b}e^{\frac{U(x)}{q}}dx
\int_{- \infty}^{x}e^{\frac{-U(y)}{q}}dy.
\label{taudef}
\end{equation}
If the starting position of the particle is between the maximum
and the right boundary of the potential $(L<x_{0}<b)$, then the
initial state is unstable. Specifically if the initial position of
the particle is between the maximum and the crossing point of the
potential with the $x$-axis $x_c = (L + E/k)$, then the average
escape time (MFPT) rises to infinity when $q\rightarrow0$, while
for zero noise we obtain a finite deterministic decay time. In
fact from Eq.~(\ref{taudef}), for $\Delta E < E$ and $q
\rightarrow 0$, we obtain

\begin{equation}
\tau(x_0,b) \simeq \frac{q}{kh}e^{\frac{(E - \Delta
E)}{q}}\longrightarrow \infty,\quad \texttt{for} \quad q
\longrightarrow 0,
\label{tau_singular}
\end{equation}
where $\Delta E = k(x_0 - L)$. The average escape time has
therefore a singularity at $q=0$ for the following range of
starting positions of the particle: $L<x_{0}<x_c$. When the
initial position is between the crossing point and the absorbing
boundary ($x_c<x_{0}<b$), then $\Delta E > E$ (see Fig. \ref
{piecewise}) and the average escape time has a nonmonotonic
behavior, with a maximum, as a function of noise intensity
\cite{Agu1}. The qualitative mechanism of this phenomenon is as
follows: a small quantity of noise can push the particle into
potential well, then the particle will be trapped there for a long
time because the well is deep with respect to the noise intensity
considered. As a consequence the NES effect appears for a fixed
potential with a metastable state if the initial position of the
particle is within the range $L<x_{0}<x_c$ \cite{Agu1}.\\
\indent We consider now the initial state at $x(0)=0$ and
\emph{dichotomous driving force} $F(t)=\pm a$ with period $T$.
When $F(t) = -a$, the initial state $x(0) = 0$ is metastable,
while for $F(t) = +a$, it becomes unstable. Potential $U(x)$ is
defined by Eq.~(\ref{f-2}) where $h=0$. We choose $F(t)=+a$ $(0\le
t<T/2)$, \ie the potential barrier is absent for the first half of
a period. In the absence of noise Eq.~(\ref{f-1}) has a periodical
solution in the deterministic regime for $T<2L/a$ and the particle
always remains trapped in the metastable state $(x(t)<L)$. This is
the overall stable regime. We consider overall unstable regime
when the period of the driving force is
\begin{equation}
T>\frac{2L}{a} \ .
\label{f-3}%
\end{equation}
In this case the particle surmounts the region $\left[ 0,L\right]
$ at time $t=L/a$ and reaches some point $x_1$, between $L$ and
boundary $b$, at time $t=T/2$, and then crosses the absorbing
boundary. If we add a small quantity of noise into the system, the
position of the particle at time $t=T/2$ is almost the same:
$x_1(q)\approx x_1(0)$. The decay time for an initial position
$x(0)=0$ is therefore $\tau(0,q)\approx T/2+\tau(x_{1},q)$, and
$\tau(x_{1},q)\gg\tau(x_{1},0)$ because the potential barrier,
which appears at $t=T/2$, makes the average escape time very large
just for $q\rightarrow0$, in accordance with
Eq.(\ref{tau_singular}). This means that the particle at time $t =
T/2$ is in an unstable position with a potential well on the left
(see Fig. \ref{piecewise}). All the trajectories that put the
particle into the metastable state contribute to increase the
average escape time $\tau$ with respect to the dynamical time,
producing a nonmonotonic behavior of $\tau$ as a function of noise
intensity. The decay time $\tau(0,q)$ will increase with small $q$
and the NES phenomenon appears. Thus, the NES effect will occur,
if the position $x$ of the particle at time $t=T/2$ is in the
following range
\begin{equation}
L<x\left(T/2\right)  <b.
\label{f-4}%
\end{equation}
This condition can be rewritten as follows \cite{Agu1}
\begin{equation}
\frac{L}{a}<\frac{T}{2}<\frac{ab+kL}{a\left( a+k\right)  }.
\label{f-5}%
\end{equation}

This inequality together with the condition $a<k$ gives the area
on the parameters region $(T,a)$ where the NES effect takes place
for small noise intensity with respect to the barrier height (see
Fig.~2).
\begin{figure}[htbp]
\begin{center}
\includegraphics*[angle=0,width=9cm]{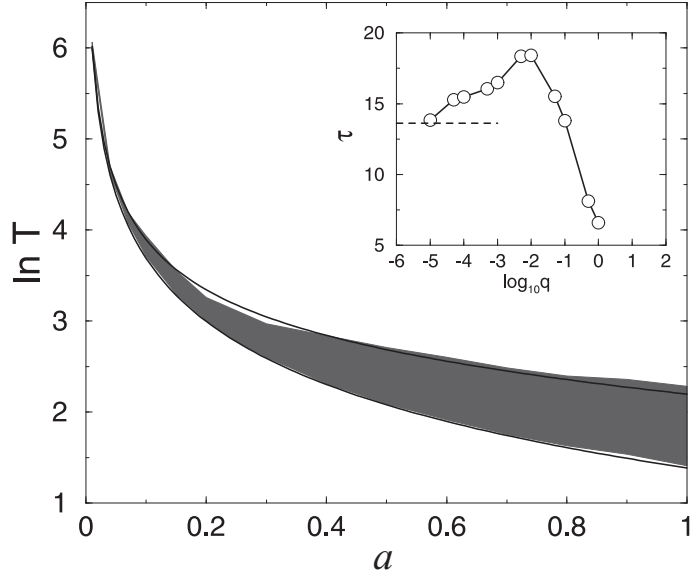}
\end{center}
\caption{\small \emph{The shaded area, obtained by numerical
simulations, is the region of the plane $(lnT,a)$ where the NES
effect appears for a dichotomous driving force. The lower and
upper continuous lines correspond to the left and to the right
sides of inequality (\ref{f-5}). The average escape time is
greater than $10\%$, above the deterministic escape time, near the
lower boundary. The parameters are: $b=7$, $k=1$, $L=2$. Inset:
the average escape time versus the noise intensity for $a=0.3$ and
$T=13.5$. The dashed line indicates the deterministic escape
time.}} \label{dicdri}
\end{figure}
Below the lower boundary we obtain Kramers-like behavior. The
magnitude of the NES effect decreases from the lower to the upper
boundary. This is because near the upper boundary the potential
barrier is very small or absent during the noise induced escape
process.

\subsection{Sinusoidal Driving}

In the case of sinusoidal force $F(t)=a\cdot\sin\omega t$ we
consider the same fixed potential profile $U(x)$ of
Eq.~(\ref{f-2}) with $h=0$. The solution in the deterministic
regime is $x(t)=(a/\omega)\cdot(1-\cos\omega t)$ for
$\omega>2a/L$. In this case of overall stable regime, $x(t)<L$ for
any $t$ and the particle always remains in the metastable state.
If the frequency is $\omega<2a/L$ the particle surmounts the
region $[0,L]$ and the solution reads
\begin{equation}
x\left(  t\right)  =\left\{
\begin{array}
[c]{lll}%
(a/\omega)\cdot(1-\cos\omega t), \qquad \qquad \quad
0<t<\theta_{1},\;0<x\left( t\right) <L,\\ k\left(
t-\theta_{1}\right) +(a/\omega)\cdot(1-\cos\omega t), \quad
t>\theta _{1},\; L<x\left( t\right) <b,
\end{array}
\right.
\label{f-6}%
\end{equation}
where $\theta_{1}$ is the time at which the particle crosses the
point $x=L$. Since the mechanism of NES effect is the same as for
dichotomous driving we can apply the same condition (\ref{f-4})
for the effect occurrence. For the sinusoidal driving this
condition can be rewritten as follows
\begin{equation}
\frac{k}{b}\left[  \pi-\arccos\left(  1-\frac{\omega L}{a}\right)
\right]+\frac{2a}{b}
<\omega<\frac{2a}{L}.\label{f-7}%
\end{equation}

This inequality and the condition $a<k$ give the area on the
parameters region $(a,\omega)$ where the NES effect takes place
for small noise intensity with respect to the barrier height (see
Fig.~3).
\begin{figure}[htbp]
\begin{center}
\includegraphics*[height=9cm,width=9cm]{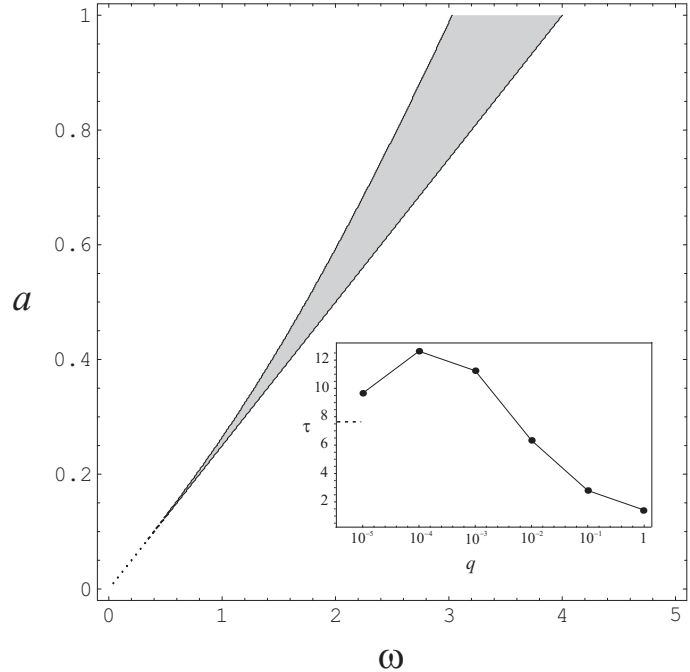}
\end{center}
\caption{\small \emph{The shaded area is the region of the plane
$(a,\omega)$ where the NES effect takes place for a sinusoidal
driving force. The parameters are: $b=1$, $k=1$, $L=0.5$. Inset:
the average escape time versus the noise intensity for $a=0.1$ and
$\omega=0.39$. The dashed line indicates the deterministic escape
time.}}
\label{sindri}
\end{figure}
So the particle, after $t = T/2$, has the potential well on the
left as in previous case (see Fig. \ref{piecewise}) and as a
result the average escape time will increase with $q$ and the NES
phenomenon takes place. When the frequency $\omega$ and the
amplitude $a$ are chosen exactly on the left hand boundary of
Eq.~(\ref{f-7}), the maximal height of induced potential barrier
is zero, and the effect is very small. When we move to the right
hand boundary of Eq.~(\ref{f-7}) the maximal barrier increases and
the NES phenomenon too. After we cross the right boundary the
deterministic decay time becomes infinite and the NES effect
disappears. Then, in the presence of noise we get Kramers-like
behaviour.

\section{Dichotomous Random Force}

\subsection{Mean First Passage Time}

We consider now a randomly switching potential profile with
reflecting boundary at $x=0$ and absorbing boundary at $x=b$. In
Eq.~(\ref{f-1}) $U(x)$ is a fixed potential and $F(t)=a\eta(t)$
where $\eta(t)$ is a Markovian dichotomous process which takes the
values $\pm 1$ with the mean rate of switchings $\nu $. Exact
results of MFPT for non-Markovian processes driven by two-state
noise and without thermal diffusion $(q=0)$ have been obtained in
ref.~\cite{Han1} and then generalized by various authors (see, for
example, \cite{Mas}). Exact equations for MFPTs for Brownian
diffusion in switching potentials were first derived in
\cite{Bal}.

From the backward Fokker-Plank equation and applying our boundary
conditions
\begin{equation}
T_{\pm }^{\prime }\left( 0\right) =0,\qquad T_{\pm }\left(
b\right) =0 \label{cond-1}
\end{equation}
we obtain the following coupled differential equations
\begin{eqnarray}
qT_{+}^{\prime \prime }+\left[ a-U^{\prime}\left(x\right) \right]
T_{+}^{\prime }+\nu \left( T_{-}-T_{+}\right) & =-1, \nonumber \\
qT_{-}^{\prime \prime }-\left[ a+U^{\prime}\left(x\right) \right]
T_{-}^{\prime }+\nu \left( T_{+}-T_{-}\right) & =-1. \label{Hang}
\end{eqnarray}
Here $T_{+}(x)$ and $T_{-}(x)$ are the mean first-passage times
for initial values $\eta (0)=+1$ and $\eta (0)=-1$ respectively.
Introducing two auxiliary functions
\begin{equation}
T=\frac{T_{+}+T_{-}}{2},\qquad \theta =\frac{T_{+}-T_{-}}{2}
\label{newT}
\end{equation}
we can write the boundary conditions (\ref{cond-1}) in the form:
$T^{\prime }(0) =\theta ^{\prime }(0) = 0$, $T(b) =\theta (b) =
0$. So that from Eqs.~(\ref{Hang}) and (\ref{newT}) we obtain
\begin{eqnarray}
qT^{\prime \prime }-U^{\prime}\left(x\right) T^{\prime }+a\theta
^{\prime
}& =&-1, \nonumber \\
q\theta ^{\prime \prime }-U^{\prime}\left(x\right) \theta ^{\prime
}+aT^{\prime }-2\nu \theta & =&0. \label{twoT}
\end{eqnarray}
After removing $T(x)$ from Eqs.~(\ref{twoT}) we obtain a
third-order linear differential equation for the variable
$\theta(x)$
\begin{equation}
\theta ^{\prime \prime \prime }-\frac{2U^{\prime}\left(x\right)
}{q}\theta ^{\prime \prime }+\left[ \frac{U^{\prime
2}\left(x\right) }{q^{2}}-\frac{U^{\prime \prime }\left(x\right)
}{q}-\gamma ^{2}\right] \theta ^{\prime }+\frac{2\nu
U^{\prime}\left(x\right) }{q^{2}}\theta =\frac{a}{q^{2}},
\label{theta}
\end{equation}
where
\begin{equation}
\gamma =\sqrt{\frac{a^{2}}{q^{2}}+\frac{2\nu }{q}}. \label{gamma}
\end{equation}
We analyze here the piecewise linear potential (\ref{f-2}) with
$h=0$. We have a metastable state for $\eta (t)=-1$ and an
unstable state for $\eta (t)=+1$. Let us focus on $T_{+}(0)$
corresponding to a finite deterministic escape time. We solve
Eqs.~(\ref{twoT}) and (\ref{theta}) separately for regions $0\le
x\le L$ and $L\le x\le b$. Using the boundary conditions and the
continuity conditions at the point $x=L$, we obtain for small
noise intensity
\begin{equation}
T_{+}\left( q\right) \simeq T_{+}\left( 0\right)+ \frac{q}{a^{2}}
\cdot f(\beta,\omega,s) + o(q), \label{main}
\end{equation}
where
\begin{eqnarray}
f(\beta,\omega,s) &=&
\frac{\beta ^{3}\left[ 2+s(1+\beta ^{2})\right] }{%
\left( 1+\beta \right) \left( 1-\beta ^{2}\right)
}\e^{-s}+\frac{\beta \left( 1-\beta ^{2}-2\beta ^{3}\right)
}{2\left( 1-\beta ^{2}\right) }\left( 1-\e^{-s}\right) \nonumber \\
& & -\frac{5+\beta }{2\left( 1+\beta \right) }+2\omega \left(
\frac{1}{1-\beta ^{2}}-\frac{3}{\beta }\right) -\frac{2\omega
^{2}}{\beta ^{2}} \label{effebeta}
\end{eqnarray}
with dimensionless parameters $\beta $, $\omega $ and $s$
\begin{equation}
\beta =\frac{a}{k},\qquad \omega =\frac{\nu L}{k},\qquad
s=\frac{2\nu \left(
b-L\right) }{k\left( 1-\beta ^{2}\right) }=\frac{2\omega }{1-\beta ^{2}}%
\left( \frac{b}{L}-1\right), \label{d-less}
\end{equation}
and
\begin{equation}
T_{+}\left( 0\right) =\frac{2L}{a}+\frac{\nu L^{2}}{a^{2}}+\frac{b-L}{k}%
\left[ 1-\frac{\beta }{s\left( 1+\beta \right) }\left(
1-\e^{-s}\right) \right] \label{zero}
\end{equation}
is the MFPT $T_{+}(0)$ of initially unstable state in the absence
of thermal diffusion.

The condition to observe the NES effect can be expressed by the
following inequality
\begin{equation}
f(\beta,\omega,s)>0. \label{NES}
\end{equation}
Let us analyze the structure of NES phenomenon region on the plane
$(\beta ,\omega )$. At very slow switching $\nu \rightarrow 0$
$(\omega \rightarrow 0,\;s\rightarrow 0)$, in accordance with
Eq.~(\ref{effebeta}), the inequality (\ref{NES}) takes the form
\begin{equation}
\beta >0,802;\quad \omega <\frac{2\beta ^{2}\left( 1-\beta \right)
-5\beta \left( 1-\beta ^{2}\right) ^{2}/2}{6\left( 1-\beta
^{2}\right) ^{2}-2\beta \left( 1-\beta ^{2}\right) +\beta
^{2}\left( 3\beta ^{2}-1\right) \left( b/L-1\right) }.
\label{regions}
\end{equation}
In the case of $\beta \simeq 1$ we obtain from
Eqs.~(\ref{effebeta}),(\ref{NES}) and (\ref{regions})
\begin{equation}
\omega <\frac{1-\beta }{b/L-1},\qquad
\frac{1}{2}+\frac{5}{2}\left( 1-\beta \right) <\omega
<\frac{1}{2(1-\beta )}.
\end{equation}

In Fig.~4 are shown two NES regions (shaded area) on the plane
$(\beta ,\omega )$.
\begin{figure}[htbp]
\begin{center}
\includegraphics*[height=6.5cm,width=6.5cm]{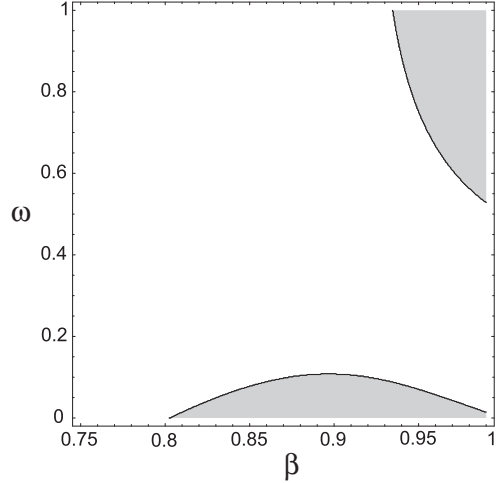}
\end{center}
\caption{\small \emph{The shaded areas are the region of the plane
$(\omega$,$\beta )$ where the NES effect takes place. The
parameters are: $L=1$, $b=1.2$, $k=1$.}} \label{nesmfpt}
\end{figure}
The NES effect occurs at the values of $\beta \simeq 1$, \ie at
very small steepness $k-a=k(1-\beta )$ of the reverse potential
barrier for the metastable state. For such a potential profile, a
small noise intensity can move back Brownian particles into
potential well, after they crossed the point $x=L$, increasing the
MFPT.

\subsection{Nonlinear Relaxation Time}

Now we consider the nonlinear relaxation time (NLRT) for the
system with randomly fluctuating potential of the previous
paragraph. The NLRT implies to take into account the inverse
probability current through the boundary or equivalently to
consider the absorbing boundary at infinity \cite{Agu3}. By
considering the well-known expression for probability density of
the process $x(t)$
\[
W\left(x,t\right) =\left\langle
\delta\left(x-x\left(t\right)\right)\right\rangle
\]
and using the auxiliary function
\[
Q\left(  x,t\right) = W\left( x,t\right) \left\langle \eta\left(
t\right) \mid x\left( t\right) =x\right\rangle
\]
we obtain the following set of closed equations for the functions
$W\left(x,t\right)$ and $Q\left( x,t\right)$ \cite{Agu2,Luc}
\begin{eqnarray}
\frac{\partial W}{\partial t} =\frac{\partial}{\partial x}\left[
U^{\prime}\left(  x\right)  W\right]  -a\frac{\partial Q}{\partial
x}+q\frac{\partial^{2}W}{\partial x^{2}%
},\nonumber\\
\frac{\partial Q}{\partial t} =\frac{\partial}{\partial x}\left[
U^{\prime}\left(  x\right)  Q\right]  -a\frac{\partial W}{\partial
x}+q\frac{\partial^{2}Q}{\partial x^{2}%
}-2\nu Q.\label{f-8}
\end{eqnarray}
The initial conditions for these functions are:
$W(x,0)=\delta(x-x_{0}),\;Q(x,0)=0$. We consider potential
profiles $U(x)\pm ax$ with a reflecting boundary at $x=0$ and an
absorbing boundary at $x\rightarrow+\infty$. We assume that the
potential profile $U(x)+ax$ corresponds to a metastable state and
$U(x)-ax$ corresponds to an unstable state. The average escape
time from metastable state within the interval $(L_{1},L_{2})$ is
defined as follows
\begin{equation}
\tau\left(
x_0\right)=\int_{0}^{+\infty}dt\int_{L_{1}}^{L_{2}}W\left(
x,t\left\vert x_0,0\right. \right)
dx.\label{f-9}%
\end{equation}
To obtain the escape time we generalize the method, proposed in
\cite{Mal1}, for fluctuating potentials. As it is shown in that
work the escape time (\ref{f-9}) can be expressed in terms of the
function $Z_{1}(x,x_0)$
\[
\tau (x_0)=\int_{L_{1}}^{L_{2}}Z_{1}\left(  x,x_0\right) dx,
\]
where $Z_{1}(x,x_0)$ is the linear coefficient of the expansion of
the function $sY(x,x_0,s)$ in a power series in $s$ and
$Y(x,x_0,s)$ is the Laplace transform of the conditional
probability density $W(x,t\left\vert x_0,0\right. )$. By Laplace
transforming the auxiliary function $Q(x,t)$ in $R(x,x_0,s)$ and
expanding the function $sR(x,x_0,s)$ in similar power series we
obtain from Eq.~(\ref{f-8}) and the conditions of zeroth
probabilistic flow at reflecting boundary $x=0$
\[
\left[ aQ-U^{\prime}\left( x\right) W-qW^{\prime}\right] _{x=0}=0,
\quad \left[ aW-U^{\prime}\left( x\right) Q-qQ^{\prime}\right]
_{x=0}=0
\]
the following coupled integro-differential equations for the
functions $Z_{1}(x,x_0)$ and $R_{1}(x,x_0)$
\begin{eqnarray}
qZ_{1}^{\prime}+U^{\prime}\left(  x\right)
Z_{1}-aR_{1} =-1\left( x-x_{0}\right) ,\nonumber\\
qR_{1}^{\prime}+U^{\prime}\left(  x\right) R_{1}-aZ_{1} =2\nu
\int_{0}^{x}R_{1}(y,x_0)dy, \label{f-10}
\end{eqnarray}
where $R_{1}(x,x_0)$ is the linear coefficient of the expansion of
the function $sR(x,x_0,s)$ and $1(x)$ is the step function. We
consider now the same piecewise linear potential profile
(\ref{f-2}) with $h=0$ and $b\rightarrow +\infty $. The initial
position of the Brownian particles is $x_{0}=0$. Solving the set
of Eqs.~(\ref{f-10}) with the continuity conditions at the point
$x=L$, we obtain the final expression for the lifetime of
metastable state $\left( L_{1}=0, L_{2}=L\right)$
\begin{equation}
\tau (0)=c_{1}\left(  \frac{\sinh\gamma L}{\gamma L}+\frac{2\nu
q}{a^{2}}\right) +c_{2}\left(  \cosh\gamma L-1\right) -\frac{\nu L^{2}}%
{q^{2}\gamma^{2}}, \label{f-11}
\end{equation}
where $c_{1},c_{2}$ have complicated expressions in terms of the
system parameters, and $\gamma $ is given by Eq.~(\ref{gamma}).
The exact formula (\ref{f-11}) was derived without any assumptions
on the thermal noise intensity $q$ and the mean rate of switchings
$\nu$. From Eq.~(\ref{f-11}) we obtain explicit expressions of the
asymptotic behaviors of the average escape time as a function of
the noise intensity $q$ and the system parameters. Specifically
for $q\rightarrow\infty$ we find
\begin{equation}
\tau (0)=\frac{L}{k}+\frac{L^{2}}{2q}\left(  1+\frac{a^2}{\nu
kL}\right) +o\left(  q^{-1}\right)  .
\label{f-11b}
\end{equation}
Thus, the average escape time decreases with $q$ and tends to a
constant value $L/k$ at $q\rightarrow\infty$. For very high noise
intensity the Brownian particle "does not see" the fluctuations of
the potential and moves as in a fixed potential profile:
$U(x)=-kx$. In the opposite limiting case of very slow diffusion
$(q\rightarrow0)$, using truncated expansions and algebraic
manipulations we obtain
\begin{equation}
\tau (0)= \tau_d + \frac{q}{a^{2}%
}\left[  \frac{a\left(  2k-a\right)  }{k^{2}-a^{2}}-3+\frac{2\nu
L}{a}\left( \frac{ka}{k^{2}-a^{2}}-3\right)
-\frac{2\nu^{2}L^{2}}{a^{2}}\right] +o\left(  q\right),
\label{f-12}
\end{equation}
where
\[
\tau_d = \frac{\nu L^{2}}{a^{2}}+\frac{1}{2\nu}+\frac{2L}{a}.
\]
In the absence of thermal diffusion $(q=0)$, at the limiting cases
$\nu\rightarrow0$ and $\nu\rightarrow\infty$ the average escape
time becomes infinite: $\tau_d\rightarrow\infty$. For
$\nu\rightarrow0$ the metastable state becomes stable and
therefore is long-lived. For $\nu\rightarrow\infty$ the switchings
are so fast that Brownian particles remain practically in the
initial point $x_{0}=0$. The lifetime is minimum when the mean
rate of switchings is equal to $a/(L\sqrt{2})$. To obtain NES
effect in the system investigated the term in quadratic brackets
in Eq.~(\ref{f-12}) must be positive. Introducing the same
dimensionless parameters $\beta $ and $\omega $ (see
Eq.~(\ref{d-less})) we can write the condition for the NES
phenomenon in the form of inequality
\begin{equation}
\omega<\frac{\beta}{2}\left[  \sqrt{\frac{1}{\left( 1-\beta
^{2}\right) ^{2}}-\frac{2\beta +3}{1-\beta
^{2}}+5}+\frac{\beta}{1-\beta^{2}}-3\right]
,\quad\beta>\frac{\sqrt{7}-1}{2}.\label{f-18}%
\end{equation}
The NES effect occurs mainly at the values of $\beta$ near $1$,
\ie at very small steepness $k-a=k(1-\beta)$ of the reverse
potential barrier beyond the metastable state as in the case of
MFPT (see Fig.~5).
\begin{figure}[htbp]
\begin{center}
\includegraphics*[height=6.5cm,width=6.5cm]{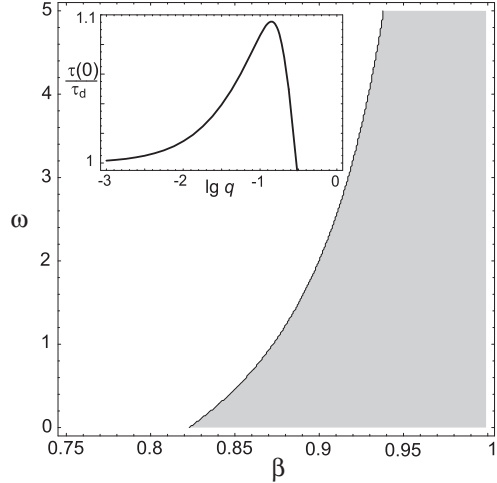}
\caption{\small \emph{The shaded area is the parameter region on
the plane $(\beta ,\omega )$ where the NES effect can be observed.
The parameters are: $L = 1$, $k = 1$. Inset: the average escape
time versus the noise intensity for $\beta = 0.97$ and $\omega =
0.1$.}} \label{nes2}
\end{center}%
\end{figure}

Only Brownian particles that are put back into the potential well
by a very small thermal noise intensity produce NES phenomenon. In
Fig.~5 we plot in the inset the normalized average lifetime as a
function of the noise intensity for $\omega = 0.1$ and $\beta =
0.97$.

\section{Conclusion}

In this paper we have presented a short review of noise enhanced
stability (NES) phenomenon. After shortly reviewing several
physical systems where the nonmonotonic behavior or resonance-like
phenomenon of the average escape time as a function of the noise
intensity was observed, we have presented the theoretical
approaches that we used to explain the NES effect for systems with
periodically driven or fluctuating metastable state. The
variations of the potential are due to: (i) a periodical force,
(ii) a Markovian dichotomous noise. For periodical driving force
we obtained the conditions and the parameter region where the NES
effect can be observed. Using the backward Fokker-Planck equation
and the Laplace-transform method we obtained the exact expressions
of the MFPT and NLRT for randomly fluctuating metastable state in
piecewise linear potential profile with reflecting boundary at the
origin. These expressions are valid for arbitrary noise intensity
and for arbitrary fluctuation rate of the potential. The analysis
at small thermal noise intensity allowed us to obtain analytically
the region of NES phenomenon occurrence in case (ii). In contrast
with the case of periodically driven metastable state, in the
presence of a random dichotomous noise the NES effect can be
observed only at very flattened potential profile beyond the
potential well, \ie in the absence of the reverse potential
barrier for particles beyond the metastable state. Only in such a
situation Brownian particles which are at large distances from the
origin can turn back into potential well by low noise intensity,
producing an enhancement of stability of
the metastable state of the system.\\
\indent All the nonmonotonic behaviors observed in different
physical systems and related with the NES effect allows us to
conclude that this  phenomenon provides a quite unexpected way to
enhance the stability of metastable states.

\section{Acknowledgments}
This work has been supported by INTAS Grant 2001-0450, MIUR, INFM,
by Russian Foundation for Basic Research (project 02-02-17517), by
Federal Program "Scientific Schools of Russia" (project
1729.2003.2), and by Scientific Program "Universities of Russia"
(project 01.01.020).


\begin{thebibliography}{99}

\bibitem{Kra}H.A. Kramers, Physica \textbf{7}, 284 (1940).

\bibitem{Han}P. H\"{a}nggi, P. Talkner, M. Borkovec,
\textit{Rev. Mod. Phys.} \textbf{62}, 251 (1990); V.I. Mel'nikov,
\textit{Phys. Rep.} \textbf{209}, 1 (1991); P. H\"{a}nggi, in
\textit{New Trends in Kramers' Reaction Rate Theory}, edited by P.
Talkner and P. H\"{a}nggi, Kluwer Acad. Pub., Dordrecht 1995,
p.93.

\bibitem{Hal}J. Hales, A. Zhukov, R. Roy, M.I. Dykman,
\textit{Phys. Rev. Lett.} \textbf{85}, 78 (2000); Sh. Kogan,
\textit{Phys. Rev. Lett.} \textbf{81}, 2986 (1998).

\bibitem{Rei}P. Reimann, \textit{Phys. Rep.} \textbf{361}, 57 (2002);
M. Muthukumar, \textit{Phys. Rev. Lett.} \textbf{86}, 3188 (2001).

\bibitem {Gam}L. Gammaitoni, P. H\"{a}nggi, P. Jung, F. Marchesoni,
\textit{Rev. Mod. Phys.} \textbf{70}, 223 (1998); R.N. Mantegna,
B. Spagnolo, \textit{Phys. Rev.} \textbf{E49}, R1792 (1994); R.N.
Mantegna, B. Spagnolo, M. Trapanese, \textit{Phys. Rev.}
\textbf{E63}, 011101 (2001); E. Lanzara, R.N. Mantegna, B.
Spagnolo, R. Zangara, \textit{Am. J. Phys.} \textbf{65}, 341
(1997).

\bibitem{Sme}V.N. Smelyanskiy, M.I. Dykman, B. Golding, \textit{Phys.
Rev. Lett.} \textbf{82}, 3193 (1999); M. Array\'as, M.I. Dykman,
R. Mannella, P.V.E. McClintock, N.D. Stein, \textit{Phys. Rev.
Lett.} \textbf{84}, 5470 (2000).

\bibitem {Leh}J. Lehmann, P. Reimann, P. H\"{a}nggi, \textit{Phys. Rev.
Lett.} \textbf{84}, 1639 (2000); \textit{Phys. Rev.} \textbf{E62},
6282 (2000); R.S. Maier, D.L. Stein, \textit{Phys. Rev. Lett.}
\textbf{86}, 3942 (2001).

\bibitem {Agu1}N.V. Agudov, B. Spagnolo, \textit{Phys. Rev.}
\textbf{E64}, 035102(R) (2001); in \emph{Stochastic and Chaotic
Dynamics in the Lakes}, edited by D.S. Broomhead et al., AIP,
Melville, New York 2000, Vol. 502, p. 272.

\bibitem{Fia}A. Fiasconaro, D. Valenti, B. Spagnolo, \textit{Physica}
\textbf{A325}, 136 (2003); \textit{Modern Problems of Statistical
Physics} \textbf{2}, 101 (2003).

\bibitem{Agu2}N.V. Agudov, A.A. Dubkov, B. Spagnolo, \textit{Physica}
\textbf{A325}, 144 (2003).

\bibitem {Day}I. Dayan, M. Gitterman, G.H. Weiss, \textit{Phys. Rev.}
\textbf{A46}, 757 (1992).

\bibitem {Man}R.N. Mantegna, B. Spagnolo, \textit{Phys. Rev. Lett.}
\textbf{76}, 563 (1996); \textit{Int. J. Bifurcation and Chaos}
\textbf{8}, 783 (1998); in \emph{Stochastic Processes in Physics,
Chemistry and Biology}, Lecture Notes in Physics, edited by I.A.
Freund and T. Poeschel, Springer-Verlag, Berlin 2000, Vol. 557, p.
327; B. Spagnolo, in \emph{The Scientific and Phylosophical
Challenge of Complexity}, edited by F.T. Arecchi and M. Berti,
Fondazione Rui, Roma 2000, p.82.

\bibitem{Agu3}N.V. Agudov, \textit{Phys. Rev.} \textbf{E57}, 2618
(1998).

\bibitem{Agu4}N.V. Agudov, A.N. Malakhov, \textit{Phys. Rev.} \textbf{E60},
6333 (1999).

\bibitem{Apo}F. Apostolico, L. Gammaitoni, F.
Marchesoni, S. Santucci, \textit{Phys. Rev.} \textbf{E55}, 36
(1997).

\bibitem{Dan}D. Dan, M.C. Mahato, A.M. Jayannavar, \textit{Phys. Rev.}
\textbf{E60}, 6421 (1999).

\bibitem{Wac}R. Wackerbauer, \textit{Phys. Rev.} \textbf{E58}, 3036 (1998);
\textit{Phys. Rev.} \textbf{E59}, 2872 (1999).

\bibitem{Mie}A. Mielke, \textit{Phys. Rev. Lett.} \textbf{84}, 818 (2000).

\bibitem{Hir}J.E. Hirsch, B.A. Huberman, D.J. Scalapino, \textit{Phys. Rev.}
\textbf{A25}, 519 (1982).

\bibitem{Mal}A.N. Malakhov, A.L. Pankratov, \textit{Physica}
\textbf{C269}, 46 (1996).

\bibitem{Mah}M.C. Mahato, A.M. Jayannavar,
\textit{Mod. Phys. Lett.} \textbf{B11}, 815 (1997);
\textit{Physica} \textbf{A248}, 138 (1998).

\bibitem{Yos}M. Yoshimoto, S. Kurosawa, H. Nagashima,
\textit{J. Phys. Soc. Jpn.} \textbf{67}, 1924 (1998); K.
Matsumoto, I. Tsuda, \textit{J. Stat. Phys.} \textbf{31}, 87
(1983); M. Yoshimoto, \textit{Phys. Lett.} \textbf{A312}, 59
(2003).

\bibitem{Xie}Xie Chong-Wei, Mei Dong-Cheng,
\textit{Chin. Phys. Lett.} \textbf{20}, 813 (2003).

\bibitem{Col}P. Colet, F. De Pasquale, M. San Miguel, \textit{Phys. Rev.}
\textbf{A43}, 5296 (1991); M. San Miguel, R. Toral, in
\emph{Instabilities and Nonequilibrium Structures}, edited by
Tirapegni and W. Zeller, Kluwer Acad. Pub., Dordrecht 1997,
Vol.VI.

\bibitem{Agu5}N.V. Agudov, A.N. Malakhov, \textit{Int. J. Bifurcation and Chaos}
\textbf{5}, 531 (1995).

\bibitem{Cru}J.P. Crutchfield, J.D. Farmer, B.A. Huberman,
\textit{Phys. Rep.} \textbf{92}, 45 (1982); F. Haake, J.W. Haus,
R. Glauber, \textit{Phys Rev.} \textbf{A23}, 3255 (1981); F.T.
Arecchi, A. Politi, L. Ulivi, \textit{Nuovo Cimento} \textbf{B71},
119 (1982); P. Colet, M. San Miguel, J. Casademunt, J.M. Sancho,
\textit{Phys. Rev.} \textbf{A39}, 149 (1989).

\bibitem{Han1}P. H\"{a}nggi, P. Talkner, \textit{Phys. Rev.} \textbf{A32},
R1934 (1985).

\bibitem{Mas}J. Masoliver, K. Lindenberg, B.J. West, \textit{Phys. Rev.}
\textbf{A33}, 2177 (1986); M.A. Rodrigues, L. Pesquera,
\textit{ibid.} \textbf{A34}, 4532 (1986); C.R. Doering,
\textit{ibid.} \textbf{A35}, 3166 (1987).

\bibitem{Bal}V. Balakrishnan, C. Van den Broeck,
P. H\"{a}nggi, \textit{Phys. Rev.} \textbf{A38}, 4213 (1988).

\bibitem{Luc} J. \L uczka, M. Niemiec, E. Piotrowski, \emph{Phys.
Lett.} \textbf{A167}, 475 (1992).

\bibitem {Mal1}A.N. Malakhov, \textit{Chaos} \textbf{7}, 488 (1997).

\end{thebibliography}
\end{document}